\documentclass[12pt,preprint]{aastex}

\newcommand\bb[1] {   \mbox{\boldmath{$#1$}}  }

\newcommand\BV{Brunt-V\"{a}is\"{a}l\"{a} }

\newcommand\kva{ \bb{k\cdot v_A}  }

\newcommand\dd{\partial}
  
\newcommand\eg{{\it e.g.}}

\newcommand\ie{{\it i.e.}}
\newcommand\beq{ \begin{equation} }
\newcommand\eeq{ \end{equation} }
\newcommand\wtilde{\widetilde}

\newcommand\rsun { \,\mathrm{R_{\sun}} }

\begin{document}

\title{The Origin of Solar Activity in the Tachocline}
\author{Kyle P. Parfrey and Kristen Menou}
\affil{Department of Astronomy, Columbia University,}
\affil{550 West 120th Street, New York, NY 10027, U.S.A.}

\begin{abstract}
Solar active regions, produced by the emergence of tubes of strong
magnetic field in the photosphere, are restricted to within $35\degr$
of the solar equator. The nature of the dynamo processes that create
and renew these fields, and are therefore responsible for solar
magnetic phenomena, are not well understood. We analyze the
magneto-rotational stability of the solar tachocline for general field
geometry. This thin region of strong radial and latitudinal
differential rotation, between the radiative and convective zones, is
unstable at latitudes above $37\degr$, yet is stable closer to the
equator. We propose that small-scale magneto-rotational turbulence
prevents coherent magnetic dynamo action in the tachocline except in
the vicinity of the equator, thus explaining the latitudinal
restriction of active regions. Tying the magnetic dynamo to the
tachocline elucidates the physical conditions and processes relevant
to solar magnetism.

\end{abstract}

\keywords{hydrodynamics --- instabilities --- MHD --- Sun: activity
  --- Sun: magnetic fields --- Sun: rotation}

%\maketitle

\section{INTRODUCTION}
Magnetic fields are a leading actor throughout a star's life---from
accreting protostar to degenerate dwarf, main sequence star to
magnetar (Mestel 1999). They shape jets in some supernovae and drive a
pulsar's radiative engine. Yet we still lack a coherent theory of how
even our own star generates and sustains its field, and so are
ignorant of the wellspring of much of this evolutionary richness
(Parker 1979). Since the 1950s, much effort has been expended on
models which locate the dynamo
in the convective zone. There are several known difficulties with this
scenario (\eg, Parker 1975; Petrovay 2000). Recently, attention has
turned to the lower convective zone and the tachocline, its boundary
with the stably stratified radiative zone, in part because
stratification can anchor the strong emerging fields (Parker 1975;
Petrovay 2000; Dikpati \& Gilman 2005; Browning et al. 2006; Hughes,
Rosner, \& Weiss 2007). Dynamical considerations suggest the presence
of a large-scale magnetic field in the upper radiative zone (Gough \&
McIntyre 1998), of plausible strength $\sim 1\; \mathrm{G}$.

Here, we propose that the solar magnetic dynamo is intimately tied to the
tachocline, where latitudinal differential rotation amplifies the
toroidal field and, together with radial differential rotation, controls the
limiting role of the magneto-rotational instability (MRI). This
instability creates small-scale turbulence which tends to prevent
large-scale, ordered field growth, that leads to surface emergence and
sunspots.

The hydrodynamic (\eg, Charbonneau, Dikpati \& Gilman 1999; Dikpati \&
Gilman 2001; Arlt, Sule \& R\"{u}diger 2005) and magneto-hydrodynamic
(\eg, Gilman \& Fox 1997, Miesch, Gilman \& Dikpati 2006; Miesch 2007)
stability of the tachocline have been the subject of extensive
study. Here, we consider stability with respect to the diffusive MRI,
which has not been previously examined. Naively, one would expect that
since $\dd \Omega/\dd\theta > 0$ the tachocline would be stable to the
MRI, in the presence of the strong radial entropy stratification that
suppresses fluid motions in the radial direction. However, we show
that the inclusion of thermal diffusion enables the radial movement of
fluid on small scales, allowing the negative radial gradient of
angular velocity to be tapped, which results in instability at large
heliolatitudes.

\section{STABILITY ANALYSIS}

The tachocline is a boundary layer imposed on the solid-body-rotating
radiative zone by the convective zone, whose global angular momentum
is extracted by the magnetic solar wind (\eg, Spiegel \& Zahn 1992;
Hughes et al. 2007). Differential rotation in the tachocline, created
by external convective torques, is not necessarily reduced by
instabilities operating within it. The
rotational gradients result from the different means of energy and
momentum transport in the radiative and convective zones, so that a
tachoclinic dynamo could in principle feed on the central nuclear
energy source. Using helioseismic constraints (Thompson et al. 2003),
we consider a thin shell confined to the tachocline with differential
rotation given by an angular velocity profile
\beq
\Omega(r,\theta) = \Omega_{rad} + \frac{r - r_t + w}{2w}(1 -
\alpha_2\cos^2\theta - \alpha_4\cos^4\theta)\,\delta\Omega_{eq},
\label{omprof}
\eeq
where $\Omega_{rad}=2.69\times 10^{-6}\;\mathrm{rad\; s^{-1}}$ is
the angular velocity of the radiative zone, $\delta\Omega_{eq} =
1.08\times 10^{-7}\;\mathrm{rad\; s^{-1}}$ is the differential in
angular velocity across the tachocline at the equator, $r_t = 0.7
\rsun$ is the
radius of the tachocline's mid-point, and $2w = 0.02 \rsun$ is its
width (Basu \& Antia 2001). We use both spherical co-ordinates
$\left(r,\phi,\theta\right)$, where $\theta$ is the co-latitudinal
angle measured from the pole, and cylindrical co-ordinates
$\left(R,\phi,Z\right)$ in the following analysis. This model implies
$q\equiv \dd\ln \Omega/\dd\ln r = -10.2$ near the pole, $q = +1.4$ at
the equator and $q=0$ at $\theta = 62\degr$. Calculations are
performed at $r = r_t$. The constants $\alpha_2 = 3.56$ and $\alpha_4
= 4.21$ are derived from GONG data (Basu \& Antia 2001).

We do not have direct knowledge of the radial structure of the entropy
stratification in the tachocline, because it is not resolved by
helioseismology. Using density and pressure profiles from a standard
solar model (Bahcall, Serenelli, \& Basu 2005), we calculate an
approximate stratification profile for the tachocline, and vary the
stratification strength in our stability analysis within the range of
plausible values. We find that our results are not strongly dependent
on the assumed level of stratification.

The MRI is a local, linear, weak-field instability present in systems
with radially decreasing angular velocity (Balbus \& Hawley 1991,
1998). The most general form (Acheson 1978; Menou, Balbus \& Spruit
2004) includes the effects of entropy stratification, viscosity $\nu$,
and thermal and magnetic diffusivities ($\xi$ and $\eta$
respectively). Thermal diffusion allows a small plasma element to
radiate photons and entropically equilibrate with its surroundings
which reduces the effective stratification. Stability is determined by
the rotational gradients (including their signs) and not by the
presence of (direction-independent) shear, making the MRI
qualitatively different from shear instabilities.

Necessary and sufficient linear stability criteria for the adiabatic
MRI ($\nu = \xi = \eta =0$) can be derived via a generalized
axisymmetric Solberg-H\o iland analysis (Balbus 1995):
\begin{eqnarray}
N_r^2 + \left(r\sin^2\theta\frac{\partial}{\partial r} + \sin\theta
\cos\theta \frac{\partial}{\partial \theta}\right)\Omega^2 > 0 ,
\label{balb1}\\
N_r^2\cos\theta \frac{\partial\Omega^2}{\partial \theta} > 0,\label{balb2}
\end{eqnarray}
where $N_r$ is the radial component of the \BV (buoyancy)
frequency. In the limit of strong thermal diffusion the plasma is able
to come to entropic equilibrium with its surroundings arbitrarily
quickly, so that the stratification is effectively removed. Therefore
by setting $N_r$ to a vanishingly small constant we can use the
adiabatic criteria to 
determine magneto-rotational stability in the idealized case of
infinite thermal diffusivity. Calculated using our rotation profile
and $N_r \rightarrow 0$, Eqs. (2) \&
(3) indicate that the tachocline can be magneto-rotationally
unstable for all $\theta < 62\degr$ (Fig. 1). We vary $r_t$ and $w$
independently and find that the extent of the unstable region is
insensitive to details of the radial structure. The linear radial
profile of angular velocity assumed in our tachocline model is that
which is least magneto-rotationally unstable, because any non-linear
profile with the same angular velocities at the boundaries must have a
greater radial angular velocity gradient at some point. For example,
we find the same region of instability for an exponential profile
(Basu \& Antia 2001).

These stability criteria (with $N_r^2 \rightarrow 0$) determine the
maximum extent of the instability, because they are inclusive of every
relevant local length scale, field strength, and geometry, subject to
several assumptions (see Balbus 1995). In particular the magnetic
field must be ``weak'', \ie  having an Alfv\'{e}n
velocity less than both the rotational velocity and the sound
speed. To investigate whether instability exists for physically
reasonable scales and fields, and to confirm that it is not stabilized
by a realistic entropy stratification, we numerically solve the
triple-diffusive dispersion relation (Menou et al. 2004) for
axisymmetric modes of frequency $\omega$:
%\begin{eqnarray}
\beq
{\wtilde\omega_{b+v}}^4 {\omega_{e}}\, {k^2\over k_Z^2} + {\wtilde\omega_{b+v}}^2 {\omega_{b}}
\left[ {1\over \gamma \rho}\, \left({\cal D} P\right)\, {\cal D} \ln
P\rho^{-\gamma}\right] \nonumber\\
+ {\wtilde\omega_{b}}^2 {\omega_{e}} \left[
{1\over R^3}\, {\cal D} (R^4\Omega^2) \right] - 4 \Omega^2 (\kva)^2 {\omega_{e}}=
0,
\eeq
%\end{eqnarray}
where
\begin{eqnarray}
\bb{v_A} = { \bb{B}/\sqrt{4\pi\rho}}, \qquad
k^2 = k_R^2 + k_Z^2, {\wtilde\omega_{b+v}}^2=\omega_b \omega_v -(\kva)^2,
\qquad {\wtilde\omega_{b}}^2=\omega_b^2 -(\kva)^2, \nonumber
\end{eqnarray}
\begin{eqnarray}
\omega_b=\omega+i \eta k^2, \qquad  \omega_v=\omega+i \nu k^2,
\omega_e=\omega+ \frac{\gamma -1}{\gamma}{i  T\over  P} \chi k^2,
\qquad {\cal D} \equiv \left( \frac{k_R}{k_Z}\frac{\dd}{\dd Z}
-\frac{\dd}{\dd R}\right) . \nonumber
\end{eqnarray}

For each magnetic field configuration, we solve for two million
wavevectors distributed in $k$ phase space (guaranteeing convergence
with respect to phase space resolution) and choose the fastest growing
mode which, under exponential growth, should quickly
dominate. Following an established procedure (Menou \& Le Mer 2006), we ensure
that the modes comfortably fit within the tachocline, allowing their
vector components to be positive or negative. The density and
diffusivity values adopted are {standard:} $\rho =
0.2\;\mathrm{g\;cm^{-3}}$, $\nu = 23.6\;\mathrm{cm^2\;s^{-1}}$, $\eta
= 596\;\mathrm{cm^2\;s^{-1}}$, $\xi = 1.2\times 10^7\;\mathrm{cm^2\;s^{-1}}$.

\section{RESULTS}

We find growth rates on the order of the rotation frequency for
$|\bb{B}|=1 \;\mathrm{G}$ and $N_r^2/\Omega_{rad}^2 = 5\times 10^4$
(see Fig. 1), appropriate values for the strongly stratified lower
tachocline (\eg, Gough \& McIntyre 1998). The diffusive instability
extends to $\theta \approx 53\degr$. For this value of $N_r$, radial
fields are relatively insensitive to field strength; at
$870\;\mathrm{G}$ the growth rate peaks at a value
$\sigma_{max}/\Omega_{rad} \sim 2\times 10^{-3}$ while stabilization
occurs at $880\;\mathrm{G}$ and above. With latitudinal fields, the
growth rates decrease more rapidly as the field strength is increased:
the peak value of $\sigma_{max}/\Omega_{rad} \sim 10^{-3}$ at
$50\;\mathrm{G}$ and $\sim 10^{-6}$ at $200\;\mathrm{G}$. Instability
persists for stratifications up to and including that present at the
top of the radiative zone, $N_r^2/\Omega_{rad}^2 \approx 1.6\times
10^5$. Radial fields give higher growth rates than latitudinal ones,
with mixed geometries falling in between. As expected (Menou \& Le Mer
2006), at very small field values (\ie, $\sim 10^{-2}\;\mathrm{G}$)
diffusion causes the MRI modes to have smaller growth rates than the
hydrodynamical modes. Our survey indicates that a large region of the
parameter space is magneto-rotationally unstable for $\theta \lesssim
53\degr$, in the manner shown in Fig. 1, for field strengths from a
small fraction of a Gauss to several tens or hundreds of Gauss and for
a range of stratifications.

Diffusive instability does not extend as far in co-latitude as in the
maximally unstable {case discussed above;} the radial gradient
of angular velocity becomes less negative as the point of inflexion,
$\theta=62\degr$, is approached from the pole and its destabilizing
influence is more easily counteracted by the positive latitudinal
gradient ($\dd \Omega/\dd \theta > 0$) and stratification ($N_r^2 >
0$). For definiteness, we only show values of
$\sigma_{max}/\Omega_{rad}$ larger than $10^{-3}$. As we now discuss,
very small values may allow linear winding of the field before it is
disrupted by significant MRI exponential growth, possibly introducing
other instabilities.

Surveying relevant processes, it appears that two other weak-field
instabilities could be active in the tachocline (\eg, Acheson 1978;
Spruit 1999). Both should be secondary to the exponentially-growing
MRI when it is present with a sufficiently large growth rate, however,
since they initially rely on linear amplification of a toroidal field
by azimuthal stretching of the poloidal
component. Tayler instability is of a pinch or interchange type, to
which a toroidal field is most unstable by $m=0$ or $m=1$
perturbations (Tayler 1973). Stability is determined solely by the values
of $\dd B_{\phi}/\dd \theta$ and $\theta$ (\eg, Spruit 1999). We test the
Tayler stability of toroidal fields created by azimuthal winding of
four distinct poloidal field configurations, having poloidal
components $\bb{B}_p$ strictly along $\bb{\hat{r}}$,
$\bb{\hat{\theta}}$, $\bb{\hat{R}}$ and $\bb{\hat{Z}}$, using (Spruit 1999) 
\beq 
B_{\phi} = N\, \bb{\hat{n}} \cdot \bb{B}_p , 
\eeq where $N$
is the number of windings and $\bb{\hat{n}} \equiv
\nabla\Omega/|\nabla\Omega|$ is calculated from Eq.~(1). We find that
the instability should be largely contained within $62\degr$ or less
of the pole, although the $62\degr \lesssim \theta \lesssim 68\degr$
region could be unstable for some field geometries (Fig. 1). This
indicates a limited relevance of the Tayler process for the
tachocline, consistent with the general expectation that this process
should be active primarily in polar regions. For instance, if the
tachocline's rotational structure is related to magnetic stresses, the
weakness of $|\dd\Omega/\dd \theta|$ relative to $|\dd\Omega/\dd r|$
suggests that $B_{\theta} \gg B_r$, because the rates at which Maxwell
stresses transport angular momentum radially and latitudinally are
proportional to $B_r B_{\phi}$ and $B_{\theta} B_{\phi}$ respectively
(Spruit 2002). If so, Tayler instability would only be possible for
$\theta < 62\degr$, largely within the region where the exponentially
growing MRI dominates. Unlike the MRI, we find that the Tayler
instability depends strongly on the magnetic field geometry. In the
absence of direct constraints on $\bb{B}$ in the tachocline, we
conclude that there may be a role for the Tayler instability in the
transitional latitudes between MRI-dominance and active region
formation, $50\degr \lesssim \theta \lesssim 60\degr$.

A second weak-field instability occurs if $ p \equiv \dd \ln B/\dd \ln
r <0$, when gas is supported against gravity by magnetic
pressure. Given a sufficiently strong radial gradient, free energy can
be liberated by field-line buckling (\eg, Parker 1955; Acheson
1978). As with other weak-field instabilities, stratification (reduced by thermal
diffusion) will be stabilizing. In the tachocline, at the equator, the
instability criterion is $|p|B^2 \gtrsim 10^{11} \;\mathrm{G^2}$
(Acheson 1978; Spruit 1999); assuming a non-singular $p \sim -1$
value, diffusive buoyancy instability would require $|\bb B | \gtrsim
10^{5} \;\mathrm{G}$, which is also the field strength at which
large-scale adiabatic buoyant instability has been shown to cause flux
emergence resulting in active region formation (\eg, D'Silva \&
Choudhuri 1993; Sch\"{u}ssler et al. 1994). Therefore the small-scale
diffusive buoyant instability appears to be of little relevance.

To summarize, we infer that the tachocline is magneto-rotationally
unstable for $\theta \lesssim 53\degr$ and stable closer to the
equator. In the unstable region we expect the small-scale turbulence produced by the
diffusive MRI, which is a narrow salt-finger-type instability
(Korycansky 1991; Menou et al. 2004), to prevent the ordered growth of
magnetic field on large scales. We surmise that in the
magneto-rotationally  stable region, latitudinal differential rotation
is free to wind the poloidal field in the azimuthal direction,
creating a linearly amplified toroidal component as in the original
Babcock-Leighton dynamo (Babcock 1961; Leighton 1964). When the
toroidal field reaches a strength of order $10^{5} \;\mathrm{G}$ it is
subject to buoyant instability and emerges. The Babcock-Leighton model
is consistent with the dearth of active regions within $8\degr$ of the
equator, since $\dd \Omega/\dd \theta \rightarrow 0$ as the equator is
approached (giving no field winding), and can account for the phase
dependence of the flux-emergence latitude (Babcock 1961; Dikpati \&
Charbonneau 1999).

\section{DISCUSSION \& CONCLUSION}

Strong radial shear exists in both equatorial and polar regions of the
tachocline, yet active regions are confined to $\theta \gtrsim
55\degr$. At approximately this co-latitude the negative radial
differential rotation is just strong enough to overcome stratification
and trigger the diffusive MRI, while closer to the equator the
rotational gradient is of the wrong sign. The $\theta = 55\degr$ point
does not appear to be a critical co-latitude for any other weak-field
instability, which is simple circumstantial evidence that the
diffusive MRI is a discriminant for the observed latitudinal
cutoff. We have confirmed this expectation with a rigorous stability
theorem and a detailed numerical stability analysis.

Our study is chiefly limited by its linear character, whereby it is
not possible fully to determine the effect of the saturated MRI on the
tachocline. The instability will likely generate an effective
anisotropic ``turbulent viscosity'', which could be important in explaining the
tachocline's radial thinness. Interestingly, observational constraints
on the tachocline's width imply a latitudinal thickness variation
consistent with two separate processes in action at high and low
latitudes, as in our scenario (Basu \& Antia 2001).

The identification of the tachocline as a critical region for solar
dynamo action creates opportunities to further our understanding of
solar magnetism. A particularly interesting question concerns the
complex relationship that may exist between the tachoclinic generation
of magnetic fields, which emerge to form active regions that impact
the solar magnetic wind, and the effect that this same wind has in
modifying the differential rotation in the tachocline by applying
torque at the solar surface.

The authors thank H. Spruit for insightful discussions on solar
magnetism and the dynamo problem, and the anonymous referee for
comments that improved the manuscript.

\clearpage
\begin{figure}[h!]
\centering%
\plotone{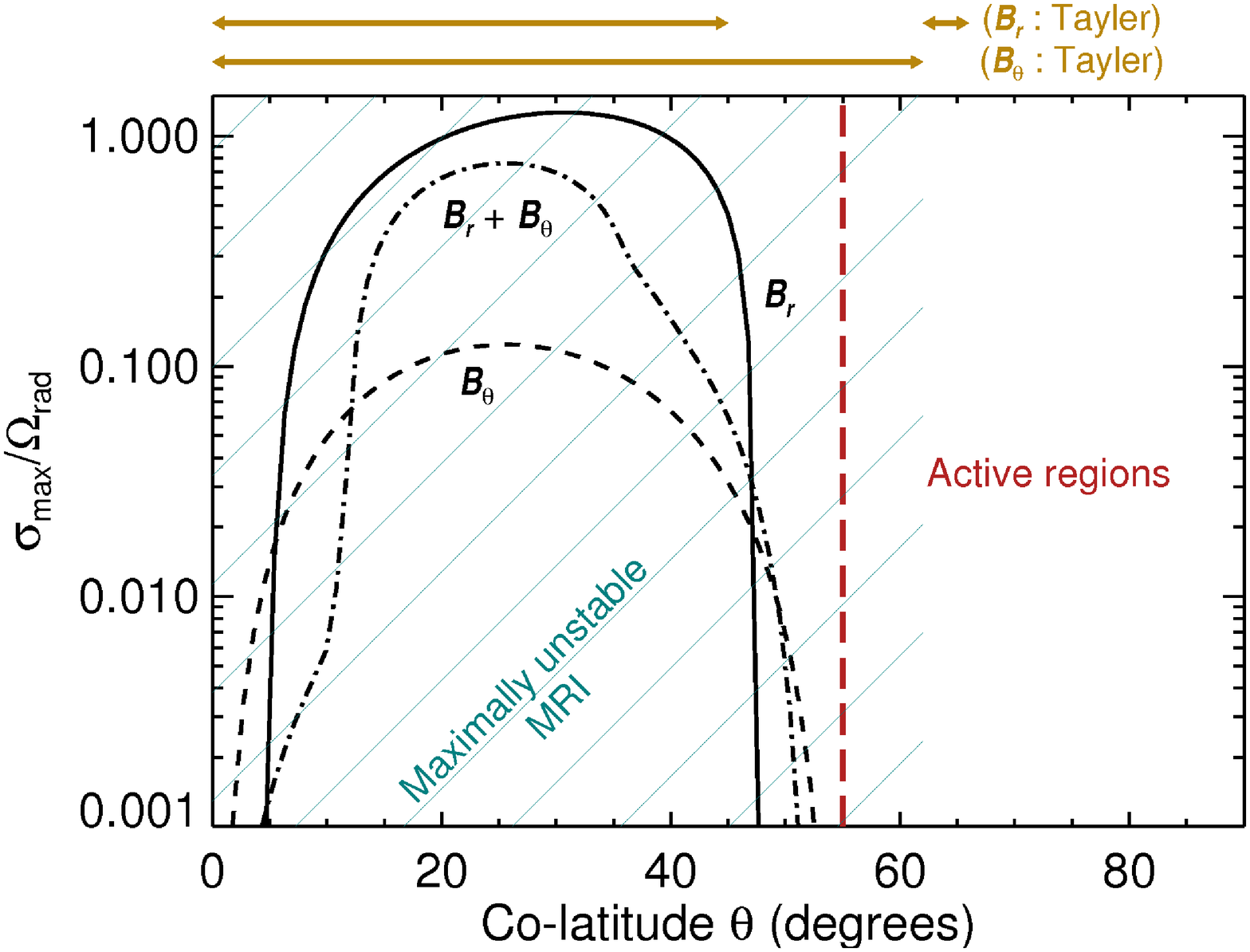}
\caption{Growth rate of the fastest growing MRI mode in units of the
rotation rate for radial ($B_r$, black solid curve), latitudinal
($B_{\theta}$, dashed), and equally radial and latitudinal fields
($B_r + B_{\theta}$, dot dashed). Values $|\bb B
|=1 \;\mathrm{G}$ and
$N_r^2/\Omega_{rad}^2=5\times 10^4$ are used in all cases. The
hatching indicates the maximally unstable region, as determined by the
Balbus stability criteria. The arrows above the plot delimit the areas
potentially subject to Tayler instability, for radial and latitudinal
fields. Active regions are found to the right of the red vertical long-dashed
line.}
\label{fig1}

\end{figure}

\end{document}